\documentclass{article}
\title{\textsc{Comparing Labeled Markov Chains: \\A Cantor-Kantorovich Approach}}
\author{Adrien Banse$^\star$\thanks{Adrien Banse is
supported by the French Community of Belgium in the framework of a FNRS/FRIA grant (email address: \texttt{adrien.banse@uclouvain.be})}~, Alessandro Abate$^\ddagger$, Raphaël M. Jungers$^\star$\thanks{RJ is a FNRS honorary Research Associate. This project has received funding from the European Research Council (ERC) under the European Union's Horizon 2020 research and innovation programme under grant agreement No 864017 - L2C, from the Horizon Europe programme under grant agreement No 101177842 - Unimaas, and from the ARC (French Community of Belgium) - project name: SIDDARTA}\\
$^\star$ICTEAM, UCLouvain\\
$^\ddagger$Department of Computer Science, University of Oxford
}
\date{}

\usepackage{amsmath}
\usepackage{amssymb}
\usepackage{color}
\usepackage{parskip}
\usepackage{cite}
\usepackage{tikz}
\usetikzlibrary{automata, positioning}

\usepackage{pdfpages}
\usepackage{fullpage}
\usepackage{amsthm}
\usepackage{bbm}
\usepackage{enumerate}
\usepackage{cite}
\usepackage{hyperref}
\usepackage{stmaryrd}

\newtheorem{theorem}{Theorem}
\newtheorem{corollary}{Corollary}
\newtheorem{lemma}{Lemma}
\newtheorem{proposition}{Proposition}

\theoremstyle{definition}
\newtheorem{definition}{Definition}

\newtheorem{example}{Example}
\newtheorem{remark}{Remark}

\makeatletter
\providecommand{\leftsquigarrow}{%
  \mathrel{\mathpalette\reflect@squig\relax}%
}
\newcommand{\reflect@squig}[2]{%
  \reflectbox{$\m@th#1\rightsquigarrow$}%
}
\makeatother

\begin{document}
	\maketitle

	\begin{abstract}
		Labeled Markov Chains (or LMCs for short) are useful mathematical objects to model complex probabilistic languages. A central challenge is to compare two LMCs, for example to assess the accuracy of an abstraction or to quantify the effect of model perturbations. In this work, we study the recently introduced Cantor-Kantorovich (or CK) distance. In particular we show that the latter can be framed as a discounted sum of finite-horizon Total Variation distances, making it an instance of discounted linear distance, but arising from the natural Cantor topology. Building on the latter observation, we analyze the properties of the CK distance along three dimensions: computational complexity, continuity properties and approximation. More precisely, we show that the exact computation of the CK distance is \#P-hard. We also provide an upper bound on the CK distance as a function of the approximation relation between the two LMCs, and show that a bounded CK distance implies a bounded error between probabilities of finite-horizon traces. Finally, we provide a computable approximation scheme, and show that the latter is also \#P-hard. Altogether, our results provide a rigorous theoretical foundation for the CK distance and  clarify its relationship with existing distances.  
	\end{abstract}

	\section{Introduction}

	\emph{Labeled Markov Chains} (or \emph{LMC}s for short) are Markov chains whose states are \emph{labeled} by symbols that belong to a finite alphabet $A$. LMCs are flexible and powerful tools to describe distributions over finite and infinite words, denoted respectively as $A^*$ and $A^\omega$, making them particularly useful in many applications ranging from computational biology \cite{Schwarz2010,CHURCHILL1989,Durbin1998,Krogh2001} to speech/gesture recognition \cite{Rabiner1989,CHEN2003745}. More recently, LMCs have been used in the context of \emph{abstraction} of continuous stochastic systems into simpler discrete model, for controller design and verification goals \cite{Banse2025,LAVAEI2022110617}. 

	More broadly, in the context of formal verification, a LMC $\tilde{\Sigma}$ is used as an \emph{approximation} of a more complex reference LMC $\Sigma$. A central question is therefore to decide whether $\Sigma$ and $\tilde{\Sigma}$ are \emph{trace-equivalent}, meaning that the two LMCs generate the same probability on $A^\omega$, the set of infinitely long words: accordingly, Larsen and Skou introduced in \cite{Larsen1991} the notion of \emph{bisimilarity}, a stronger condition than trace equivalence. However, in some applications, e.g. when $\tilde{\Sigma}$ is computed from Monte-Carlo simulations, requiring equal trace probabilities can be too restrictive. A central yet non-trivial question is therefore to \emph{quantify} the difference between two LMCs $\Sigma$ and $\tilde{\Sigma}$. 

	\begin{example}[Markov's Eugene Onegin text model \cite{markov1913example}] \label{example:biased-mc}
		In the seminal paper \cite{markov1913example}, Markov counted the transitions between vowels and consonants in one of Eugene Onegin's poem, resulting in the reference Markov chain $\Sigma$, depicted in Figure~\ref{fig:markov-example} with $\varepsilon = 0$. Suppose that such a model was computed only from a subset of the poem, then it could result in an approximate (or biased) Markov chain $\tilde{\Sigma}_{\varepsilon}$, with an $\varepsilon > 0$. Our goal is to quantify the difference between $\Sigma$ and its approximation $\tilde{\Sigma}_{\varepsilon}$.
		\qed
		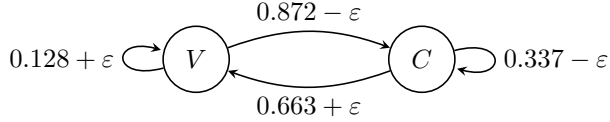
\begin{figure} 
			\centering
			\label{fig:markov-example}
			\begin{tikzpicture}[->, >=stealth, auto, node distance=3cm, semithick]
			\node[state] (V) {$V$};
			\node[state, right of=V] (C) {$C$};

			\path (V) edge[loop left] node{$0.128 + \varepsilon$} (V);
			\path (C) edge[loop right] node{$0.337-\varepsilon$} (C);

			\path (V) edge[bend left=20] node{$0.872-\varepsilon$} (C);
			\path (C) edge[bend left=20] node{$0.663+\varepsilon$} (V);
			\end{tikzpicture}
			\caption{Biased Markov's Eugene Onegin text model \cite{markov1913example}.}
		\end{figure}
	\end{example}

	Different metrics between LMCs were proposed in the literature. One can divide these metrics in two categories \cite{Jaeger2014}. The first one consists of \emph{bisimilarity metrics}, characterized by the fact that $d(\Sigma, \tilde{\Sigma}) = 0$ implies that $\Sigma$ and $\tilde{\Sigma}$ are bisimilar. In this context, Desharnais et al. introduced in \cite{Desharnais2004} the notion of \emph{$\varepsilon$-approximate bisimilarity}, naturally extending the stricter exact counterpart and devising algorithms to compute $\varepsilon$-bisimilar chains, and \cite{Bian2017} studied its relation with trace equivalence. Other examples of works studying bisimilarity distances include \cite{Breugel2007,Chen2012,vanBreugel2005}. The second category consists of \emph{trace-based distances}, directly measuring the difference between the trace probabilities, and characterized by the fact that $d(\Sigma, \tilde{\Sigma}) = 0$ implies that $\Sigma$ and $\tilde{\Sigma}$ are trace-equivalent. For example, \cite{Chen2014,Kiefer2018} consider the \emph{Total Variation} distance between the asymptotic distributions, while \cite{Daca2016,Jaeger2014} expose families of linear distances, sometimes discounted or averaged. Other examples include measures that are not metrics\footnote{In the sense that they do not necessarily satisfy symmetry, triangle inequality and/or positive-definiteness.}, e.g. \cite{Rached2004} measures the \emph{Kullback–Leibler divergence} between the asymptotic distributions. 

	Some characteristics are important to know whether a distance is well-suited for a particular application. In this paper, we focus on two types of characteristics: continuity properties and computational complexity. Notions of continuity properties for distance between LMCs were studied in \cite{Jaeger2014}. The authors highlighted two important properties, informally described hereinafter.
	\begin{itemize}
		\item Parameter continuity: a distance $d$ is \emph{parameter continuous} if, for any sequence of approximations $(\tilde{\Sigma}_n)_{n \geq 0}$ that converges to the reference $\Sigma$, it holds that $\lim_{n \to \infty} d(\tilde{\Sigma}_n, \Sigma) = 0$\footnote{Coming back to Example~\ref{example:biased-mc}, consider a sequence of Markov chains $(\tilde{\Sigma}_{\varepsilon_n})_{n \geq 0}$ corresponding to sequence $(\varepsilon_n)_{n \geq 0} \to 0$.}.
		\item $\Phi$-continuity: for a given class of properties $\Phi \subseteq 2^{A^\omega}$, e.g. LTL properties, a distance $d$ is $\Phi$-continuous if a bounded distance $d(\Sigma, \tilde{\Sigma})$ implies a bounded error when using $\tilde{\Sigma}$ instead of $\Sigma$ in all scenarii in $\Phi$. 
	\end{itemize}
	As discussed in \cite{Jaeger2014}, the two properties are conflicting with each other as the first one is trivially satisfied by $d \equiv 0$ and the other by $d \equiv 1$. In \cite{Jaeger2014}, the authors claim that a ``good distance'' is a distance that satisfies both parameter- and $\Phi$-continuity, with $\Phi$ being rather general. 

	The second property in which we are interested is the computability and the computational complexity of the distance. In particular, we are interested in knowing whether $d$ is computable and if it is hard to compute. While most distances are either not computable or hard to compute, it is also of the utmost importance to identify computable approximation schemes, and characterise their computational complexity -- see e.g. \cite{Chen2014} for a proof of NP-hardness of both the exact computation and the approximation of the asymptotic TV distance. 

	\paragraph{Contributions} Recently, the notion of \emph{Cantor-Kantorovich} distance (or \emph{CK} distance for short) was introduced in \cite{Banse2025} in the context of abstraction-based verification of dynamical systems. The goal of this paper is to extend the theoretical foundations of this distance. The main contributions are summarized below. 
	\begin{itemize}
		\item \textbf{Definition.} We make the link between the CK distance and the topology generated by the Cantor distance on $A^\omega$, making the CK distance a natural choice for comparing LMCs. We also show that the CK distance is an infinite sum of discounted TV distances, which highlights that it can be seen as an instance of the discounted linear distances studied in \cite{Jaeger2014}, but one that arises directly from the natural Cantor topology. 
		\item \textbf{Exact computation.} We show that the exact computation of the CK distance is \#P-hard. 
		\item \textbf{Continuity properties.} We first prove the continuity of the distance by showing that the distance between two similar (in a precise sense) LMCs is upper bounded. We also show that a small CK distance implies a bounded error between probabilities of finite traces and, given the distance $d(\Sigma, \tilde{\Sigma})$ and a tolerance $\varepsilon$, we provide a practical method to select  an horizon $k$ such that using $\tilde{\Sigma}$ yields an error of at most $\varepsilon$ up to that horizon $k$. Although parameter continuity and (B-LTL)-continuity were already proven in \cite{Jaeger2014} for discounted TV distances (and therefore for the CK distance), our results provide explicit quantitative bounds, alternative proofs, and practical finite-horizon guarantees.
		\item \textbf{Approximation.} We provide an approximation scheme based on finite sum of discounted TV distances. We show that this scheme is also \#P-hard. 
	\end{itemize}

	\paragraph{Outline} Section~\ref{sec:prelim} contains necessary background on distances between distributions, language theory, LMCs and computational complexity. Section~\ref{sec:ck} includes a definition of the CK distance, a link with TV distances and a proof that the exact computation of the CK distance is \#P-hard. Section~\ref{sec:continuity} contains all the results concerning the continuity properties of the CK distance. Section~\ref{sec:approx} describes an approximation scheme and a computational complexity analysis of the latter. Finally, Section~\ref{sec:conclu} gathers conclusions and open questions. 

	\paragraph{Notations} $\mathbb{R}$ and $\mathbb{N}$ respectively refer to the set of real and natural numbers, with $\mathbb{R}_{\geq 0} = \{x \in \mathbb{R} : x \geq 0\}$. Let $A, B$ be two sets. A \emph{binary relation} $R$ is a subset of $A \times B$. By abuse of notation, for any $\tilde{A} \subseteq A$, let $R(\tilde{A}) := \pi_B((\tilde{A} \times B) \cap R) \subseteq B$, where $\pi_B : A \times B \to B$ is the projection operator on $B$. And let $R(\tilde{B}) \subseteq A$ be defined similarly for any $\tilde{B} \subseteq B$. A subset $\tilde{A} \times \tilde{B}$ is said to be \emph{$R$-closed} if $R(\tilde{A}) \subseteq \tilde{B}$ and $R(\tilde{B}) \subseteq \tilde{A}$. For any set $A$, the set $A^*$ is the \emph{Kleene closure} of the set $A$, defined as $A^* := \cup_{k \geq 0} A^k$, where $A^k$ is the Cartesian $k$-th power of $A$. If $A$ is a finite alphabet, $A^\omega$ denotes the set of all infinitely long words of $A$. Let $w_1, w_2 \in A^*$ with $w_1 = (a_1, a_2, \dots)$ and $w_2 = (b_1, b_2, \dots)$, then, by abuse of notation, $w1w2 = (b_1, b_2, \dots, a_1, a_2, \dots)$. Similarly, for $W \subseteq A^\omega \cup A^*$ and $w \in A^*$, $wW := \{ww' : w' \in W\}$.  

	\section{Preliminaries} \label{sec:prelim}

	This section is divided into four parts. First, we introduce the Total Variation and Kantorovich distances between probability distributions. Second, we review some classical results from language theory and introduce the topology on the considered language, the associated Cantor distance, and LTL logic along with its fragment comprising finite-horizon B-LTL properties. Third, we formally define Labeled Markov Chains and the probability distributions they generate on the considered languages. And finally, we review some notions from computational complexity theory, including the  \#P and NP classes. 

	\subsection{Distances between Probability Distributions}

	Let $\Omega$ be any \emph{sample space}, and $\cal E$ be any \emph{event space}, where any event $E \in \cal E$ is a subspace of $\Omega$. We start by defining the \emph{Total Variation} (or \emph{TV}) distance between two probability distributions. 
	\begin{definition}[Total Variation distance]
		Let $p$ and $q$ be two probability distributions over the same pair $(\Omega, \cal E)$. The \emph{Total Variation} distance between $p$ and $q$ is given by 
		\[
			\mathsf{TV}(p, q) := \sup_{E \in \cal E} |p(E) - q(E)|. 
		\]
	\end{definition}

	The TV distance measures the maximum difference between the probabilities that $p$ and $q$ assign to the same event. If $\Omega$ is countable and $\mathcal{E} = 2^{\Omega}$, it holds that 
	\begin{equation} \label{eq:tv-computation}
		\mathsf{TV}(p, q) = \frac{1}{2} \sum_{e \in \Omega} |p(e) - q(e)|, 
	\end{equation}
	where $p(e) := p(\{e\})$ by abuse of notation, and similarly for $q(e)$. 

	We now assume that the sample space $\Omega$ is countable and that $\mathcal{E} = 2^{\Omega}$. We also assume that $\Omega$ is endowed with a distance $\mathsf{D} : \Omega \times \Omega \to \mathbb{R}_{\geq 0}$, which defines a metric space $(\Omega, \mathsf{D})$. We can then define the \emph{Kantorovich} distance\footnote{This distance is also known under other names in the literature, such as the \emph{Wasserstein}, \emph{Kantorovich-Rubinstein} or the \emph{Earth mover's} distance. See \cite{Villani2009} for an introduction.} between two probability distributions on $(\Omega, \mathsf{D})$.
	\begin{definition}[Kantorovich distance]
		Let $p$ and $q$ be two probability distributions over the same pair $((\Omega, \mathsf{D}), \cal E)$, where $(\Omega, \mathsf{D})$ is a metric space. The \emph{Kantorovich} distance between $p$ and $q$ with underlying metric $\mathsf{D}$ is given by 
		\begin{equation} \label{eq:kantorovich}
			\mathsf{K}_\mathsf{D}(p, q) := \min_{\pi \in \Pi(p, q)} \sum_{e_1, e_2 \in \Omega} \mathsf{D}(e_1, e_2) \pi(e_1, e_2), 
		\end{equation}
		where $\Pi(p, q)$ is the set of \emph{couplings} of $p$ and $q$, defined as the set of joint distributions $\pi : \Omega \times \Omega \to [0, 1]$ such that 
		\[
			\forall e_1 \in \Omega : \sum_{e_2 \in \Omega} \pi(e_1, e_2) = p(e_1), 
		\]
		\[
			\forall e_2 \in \Omega : \sum_{e_1 \in \Omega} \pi(e_1, e_2) = q(e_2). 
		\]
	\end{definition}

	One of the most direct way to gain intuition about the Kantorovich distance is to consider its \emph{optimal transport} interpretation: solving \eqref{eq:kantorovich} amounts to solving an optimal transport problem in which each resource $e \in \Omega$ has supply $p(e)$ and demand $q(e)$ (see e.g. \cite{Villani2009} for more details). 

	\subsection{Language theory} \label{sec:language}

	Let $A$ be a finite \emph{alphabet} composed of $m$ elements ($|A| = m$), the set of all words of length $k$ and all infinite words composed from alphabet $A$ are respectively denoted by $A^k$ and $A^\omega$. Let also $A^*$ be the set of all finite languages. 

	Let $w \in A^*$ be a finite word. This word defines a \emph{cylinder set} $wA^\omega \subseteq A^\omega$, defined as the set of all infinite words with prefix $w$. In this paper, we consider the \emph{natural topology} in which the cylinder sets form a basis of the open sets \cite{Jaeger2014,Perrin,Fogg2002}. This means that every element of the $\sigma$-algebra on $A^\omega$, noted $\sigma(A^\omega)$, is a union of cylinder sets.

	We consider the metric spaces $(A^*, \mathsf{C})$ and $(A^\omega, \mathsf{C})$ where $\mathsf{C}$ is the \emph{Cantor} distance, defined as follows.  
	\begin{definition}[Cantor distance] \label{def:cantor}
		For any two words $w_1, w_2 \in A^* \cup A^\omega$, their \emph{Cantor distance} is defined as 
		\begin{equation} \label{eq:cantor}
			\mathsf{C}(w_1, w_2) := m^{-\inf\{ i - 1 : a_i \neq b_i \}}, 
		\end{equation}
		where $\inf \emptyset = \infty$ by convention.
	\end{definition}
	The Cantor distance only depends on the first character that differs, and belongs to $[0, 1]$: 
	\begin{itemize}
		\item When the two words are the same ($w_1 = w_2$), $\mathsf{C}(w_1, w_2) = m^{-\inf \emptyset} = m^{-\infty} = 0$.
		\item When the two words differ from the $i$-th character with $i > 1$, $\mathsf{C}(w_1, w_2) = m^{-(i-1)} \in (0, 1)$. 
		\item When the two words differ from the very first character, $\mathsf{C}(w_1, w_2) = m^{0} = 1$.
	\end{itemize}
	Note that this choice of distance is consistent with the fact that $\sigma(A^\omega)$ is composed of all cylinder sets (see e.g. \cite{Fogg2002,Perrin,Calude2009}).

	\begin{remark} \label{rem:cantor-is-different}
		The basis of the chosen Cantor metric, that is the scalar $\ell$ such that $\mathsf{C}(w_1, w_2) = \ell^{-\inf\{ i - 1 : a_i \neq b_i \}}$ in \eqref{eq:cantor}, is arbitrary. Indeed, any choice of $\ell$ is such that the generated topology is the set of cylinder sets (see e.g. a proof in \cite{Fogg2002} for $\ell = m$, and \cite{Perrin} for $\ell = 2$). Our previous work \cite{Banse2025} considered $\ell = 2$. In this manuscript, we consider $\ell = m$ for brevity of the involved mathematical expressions. 
	\end{remark}

	In this work, we will consider traces that satisfy \emph{Bounded Linear Temporal Logic} specifications (or \emph{B-LTL} for short) \cite{Jaeger2014}. We first recall the definition of LTL specifications (see e.g. \cite{Baier2008,Belta2017}).
	\begin{definition}[LTL syntax]
		Given a finite alphabet $A$, a LTL formula $\varphi$ over $A$ is recursively defined as
		\[ 
			\varphi ::= \emph{\texttt{true}} \,|\, a \,|\, \varphi_1 \wedge \varphi_2 \,|\, \neg \varphi \,|\, \bigcirc \varphi \,|\, \varphi_1 \mathcal{U} \varphi_2, 
		\] 
		where $a \in A$, $\phi_1$ and $\phi_2$ are two LTL formulas, $\wedge$ is the \emph{conjuction}, $\lnot$ is the \emph{negation}, and $\bigcirc$ and $\mathcal{U}$ are respectively the ``next'' and the ``until'' temporal modalities. 
	\end{definition}
	A property $\phi \subset A^\omega$ is LTL definable if there exists a LTL formula $\varphi$ such that $
		\phi = \{w \in A^\omega : w \models \varphi \}, 
	$
	and the LTL class of properties $\Phi_\text{LTL} \subseteq 2^{A^\omega}$ is the set of all LTL-definable properties. 
	\begin{definition}[B-LTL syntax \cite{Jaeger2014}]
		Given a finite alphabet $A$, a B-LTL formula $\varphi$ over $A$ is recursively defined similarly as an LTL formula, however omitting the $\mathcal{U}$ temporal modality.
	\end{definition}
	A B-LTL definable property $\phi \subset A^\omega$ and the B-LTL class of properties $\Phi_{\text{B-LTL}}$ are defined similarly as above. 
	\begin{lemma}[\cite{Jaeger2014}] \label{lemma:bltl-finite-union}
		Any B-LTL definable property $\phi \subseteq A^\omega$ is a finite union of cylinder sets. 
	\end{lemma}

	\begin{remark} 
		The B-LTL class of properties is different from the \emph{sc-LTL} class of properties (\emph{syntatically co-safe LTL}, see e.g. \cite[Definition~2.3]{Belta2017}) and LTL$_f$ (LTL over finite traces \cite{Giacomo}). Indeed B-LTL specifications require a bounded lookahead (that one can encode only with ``next'' operators) whereas both sc-LTL and LTL$_f$ specifications require finite but unbounded lookahead. 
		\qed 
	\end{remark}

	\subsection{Labeled Markov Chains}

	We now define \emph{Labeled Markov Chains} (or \emph{LMC}s for short), and the probability distributions that they generate on $A^k$ and $A^\omega$. 
	\begin{definition}[Labeled Markov Chain]
		A \emph{Labeled Markov Chain} (or \emph{LMC} for short) is a tuple $\Sigma := (S, A, \mu, P, L)$ where 
		\begin{itemize}
			\item $S$ is the finite set of \emph{states}, 
			\item $A$ is the finite set of \emph{labels}, 
			\item $\mu : S \to [0, 1]$ is the \emph{initial probability distribution} and is such that \[\sum_{s \in S} \mu(s) = 1,\] 
			\item $P : S \times S \to [0, 1]$ is the \emph{transition kernel} and is such that for all $s \in S$, \[\sum_{s' \in S} P(s, s') = 1, \]
			\item $L : S \to A$ is the labeling function.
		\end{itemize}
	\end{definition}

	Similarly as in Section~\ref{sec:language}, we write $m := |A|$ to simplify notations. All along this paper, we will only consider LMC such that $m > 1$. Given an alphabet $A$ and a natural number $k \in \mathbb{N}$, a LMC $\Sigma$ naturally defines a probability distribution $p^k : A^k \to [0, 1]$ on the possible $k$-long words as follows: 
	\begin{equation} \label{eq:prob-generated}
		p^k(a_1, \dots, a_k) = \sum_{s_1 \in \llbracket a_1 \rrbracket} \mu(s_1) \sum_{s_2 \in \llbracket a_2 \rrbracket} P(s_1, s_2) \dots \sum_{s_k \in \llbracket a_k \rrbracket} P(s_{k-1}, s_k), 
	\end{equation}
	where $\llbracket a \rrbracket = \{s \in S : L(s) = a\}$. We say that the distribution $p^k$ is \emph{generated} by the LMC $\Sigma$. LMCs also naturally define an unique distribution on $\sigma(A^\omega)$ \cite{Jaeger2014}, noted $p^\omega$. 

	At the center of many research efforts is the question of deciding whether two LMCs are \emph{bisimilar} to each other (see \cite[Definition~6.1]{Larsen1991} for the original definition, and \cite[Definition~1]{Fatmi2025} for a definition in the context of LMCs). Later on, Desharnais, Laviolette and Tracol introduced in \cite{Desharnais2004} the relaxed notion of \emph{$\varepsilon$-approximately bisimilar} LMCs in order to quantify the closeness between the processes generated by the LMCs with an unique parameter $\varepsilon \in (0, 1)$ (see \cite[Definition~9]{Desharnais2008}). We formally define the notion of $\varepsilon$-bisimilarity. 

	\begin{definition}[$\varepsilon$-approximate probabilistic bisimilarity relation] \label{def:approx-bisim}
		Given two LMCs $\Sigma_1 = (S_1, A, \mu_1, P_1, L_1)$ and $\Sigma_2 = (S_2, A, \mu_2, P_2, L_2)$ defined on the same set of labels, and $\varepsilon \in (0, 1)$, an \emph{$\varepsilon$-approximate probabilistic bisimulation relation} between $\Sigma_1$ and $\Sigma_2$ is a relation $R_\varepsilon \subseteq S_1 \times S_2$ such that the following conditions hold: 
		\begin{enumerate}
			\item For all $(s_1, s_2) \in R_\varepsilon$, $L(s_1) = L(s_2)$. 
			\item For all $R_\varepsilon$-closed sets $\tilde{S_1} \times \tilde{S}_2$
			\[
			\left|\sum_{\tilde{s}_1 \in \tilde{S}_1} \mu(\tilde{s}_1) - \sum_{\tilde{s}_2 \in \tilde{S}_2} \mu(\tilde{s}_2)\right| \leq \varepsilon.
			\]
			\item For all $(s_1, s_2) \in R_\varepsilon$ and for all $R_\varepsilon$-closed sets $\tilde{S_1} \times \tilde{S}_2$,
			\[
				\left|\sum_{\tilde{s}_1 \in \tilde{S}_1} P(s_1, \tilde{s}_1)
				-
				\sum_{\tilde{s}_2 \in \tilde{S}_2} P(s_2, \tilde{s}_2)\right| 
				\leq \varepsilon.
			\]
		\end{enumerate}
	\end{definition}
	We say that two LMCs are \emph{$\varepsilon$-approximately bisimilar} to each other if there exists an $\varepsilon$-approximate probabilistic bisimulation relation between them. If $\varepsilon = 0$, then there exists a \emph{probabilistic bisimilarity relation} between the two LMCs (such as defined in \cite{Larsen1991,Fatmi2025}), and the latter are said to be \emph{bisimilar} to each other.

	\subsection{Computational complexity classes}

	In this paper, we will study the computational complexity of the exact computation of the CK distance (see Section~\ref{sec:ck}), as well as using a particular approximation scheme (see Section~\ref{sec:approx}). We therefore conclude this section by introducing the concerned computational complexity classes, and we recall some classical results from theoretical computer science. We start by introducing the class of \#P functions \cite{Bhattacharyya2023}.  

	\begin{definition}[\#P function] \label{def:sharp-P}
		A function $f : \{0, 1\}^* \to \mathbb{N}$ is in the class \#P if there is a polynomial time non-deterministic Turing machine $M$ so that for any $x$, $f(x)$ is equal to the number of accepting paths of $M(x)$. 
	\end{definition}

	The two following definitions as well as Proposition~\ref{prop:prove-hardness} are classical results from theoretical computer science. 
	\begin{definition}[\#P-hard function]
		A function $f$ is \emph{\#P-hard} if the computation of all \#P functions can be reduced to the computation of $f$ in polynomial time. 
	\end{definition}
	\begin{definition}[\#P-complete function]
		A function is \emph{\#P-complete} if it is both \#P and \#P-hard. 
	\end{definition}
	\begin{proposition} \label{prop:prove-hardness}
		If the computation of a \#P-complete function $g$ can be reduced to the computation of another function $f$ in polynomial time, then $f$ is \#P-hard. If, in addition, $f$ is \#P, then it is \#P-complete. 
	\end{proposition}

	We conclude by making the link between \#P functions and NP decision problem (see e.g. \cite{Cormen2022}). Every \#P-function has a corresponding decision problem \emph{"Given $x \in \{0, 1\}^*$, is $f(x) > 0$?"}. Since \#P is counting accepting paths while NP is deciding whether there exists at least one accepting path, one can directly see that if $f$ is \#P, then its corresponding decision problem is NP, making \#P "harder" than NP. 

	\section{The Cantor-Kantorovich distance} \label{sec:ck}

	We start by defining the \emph{Cantor-Kantorovich} (or \emph{CK}) distance between two LMCs.
	\begin{definition} \label{def:ck}
		Let $\Sigma_1$ and $\Sigma_2$ be two LMCs defined on the same set of labels. For all $k \in \mathbb{N}$, let $p_1^k$ and $p_2^k$ be the distributions generated by $\Sigma_1$ and $\Sigma_2$ respectively at step $k$. The \emph{Cantor-Kantorovich} (or \emph{CK}) distance between $\Sigma_1$ and $\Sigma_2$ is given by 
		\[
			d(\Sigma_1, \Sigma_2) := \lim_{k \to \infty} \mathsf{K}_{\mathsf{C}}(p_1^k, p_2^k). 
		\]
	\end{definition}
	This distance between LMCs is defined as the asymptotic Kantorovich distance between the generated probability distributions on the metric space $(A^\omega, \mathsf{C})$, where $\mathsf{C}$ is the Cantor distance. Because the latter generates a natural topology on $A^\omega$ (see Section~\ref{sec:language}), we claim that the CK distance is a natural distance between LMCs. 

	\begin{theorem} \label{thm:ck-is-tv}
		Let $\Sigma_1$ and $\Sigma_2$ be two LMCs defined on the same set of labels. It holds that 
		\[
			d(\Sigma_1, \Sigma_2) = \sum_{i = 1}^\infty \left(\frac{m-1}{m^i}\right) \mathsf{TV}(p_1^i, p_2^i), 
		\]
		where $p_1^i$ and $p_2^i$ are the distributions on $A^i$ generated by $\Sigma_1$ and $\Sigma_2$, respectively.  
	\end{theorem}
	For the sake of brevity, all proofs are moved to the appendices of this paper.

	\begin{remark} \label{rem:jaeger}
		Theorem~\ref{thm:ck-is-tv} is central in this paper as it provides both theoretical and practical insights about the CK distance. In particular, it shows that, up to a factor $(m-1)$, the CK distance can be seen as an instance of the discounted TV distances studied in \cite{Jaeger2014}, but one that arises directly from the natural topology generated by the underlying Cantor  metric. \qed 
	\end{remark}

	Theorem~\ref{thm:ck-is-tv} also allows to show that the CK distance is well-defined, in the sense that it converges over $k$. Note that this result was already proven in \cite{Banse2025} but in a different fashion and in a slightly different setting (see Remark~\ref{rem:cantor-is-different}). 
	\begin{corollary} \label{cor:ck-bounds}
		For any two LMCs $\Sigma_1$ and $\Sigma_2$ defined on the same set of labels, the CK distance converges and is such that $0 \leq d(\Sigma_1, \Sigma_2) \leq 1$. 
	\end{corollary}

	In the following, we provide a toy example that highlights the usefulness of the CK distance compared to other popular distances. 
	\begin{example}
		In the context of reinforcement learning~\cite{Sutton1998}, a labeled Markov chain (LMC) $\Sigma = (S, A, \mu, P, L)$ can model a \emph{Markov Decision Process} (MDP) with a given and fixed policy. Consider the following example:
		\begin{itemize}
			\item $\Sigma_{\text{exp}}$ is an LMC modeling an MDP whose policy is that of an expert (e.g., the MDP represents a vehicle, and the expert is a human driver);
			\item $\Sigma_{\text{alg}}$ is an LMC modeling an MDP whose policy was learned using a specific RL algorithm.
		\end{itemize}
		One way to measure the quality of the RL algorithm is to quantify the difference between $\Sigma_{\text{exp}}$ and $\Sigma_{\text{alg}}$. Let us consider some examples of existing distances in the litterature: 
		\begin{itemize}
			\item The \emph{Total Variation distance} between probabilities on $A^\omega$ \cite{Chen2014,Kiefer2018}, defined as $\lim_{k \to \infty} \mathsf{TV}(p_{\text{exp}}^k, p_{\text{alg}}^k)$, is overly conservative and can be equal to one even for arbitrarily close LMCs\footnote{In the sense that the two LMCs are $\varepsilon$-approximately bisimilar for arbitrarily small $\varepsilon > 0$. This is due to the fact that the TV distance defined on $A^\omega$ is not \emph{parameter continuous} (see Section~\ref{sec:continuity} and \cite{Jaeger2014}).} \cite{Daca2016}. In the context described above, this can lead to a persistent underestimation of the algorithm's performance;
			\item If both LMCs admit stationary distributions $\pi_{\text{exp}}$ and $\pi_{\text{alg}}$ on $A$, any distance $d(\pi_{\text{exp}}, \pi_{\text{alg}})$ ignores transient behaviors. In the above context, this means that an algorithm performing poorly during the transient phase, but eventually converging to the same stationary distribution, would still be considered high quality, which is misleading;
			\item Any bisimilarity metric $d(\Sigma_1, \Sigma_2)$ (e.g., \cite{Breugel2007,Chen2012,Desharnais2004,vanBreugel2005}) can be strictly positive for two LMCs that are trace equivalent but not bisimilar, thus penalizing such systems. 
		\end{itemize}
		Because the CK distance is a trace-based distance and captures differences in transient behaviors (as highlighted in Theorem~\ref{thm:ck-is-tv}), we argue that it is a particularly suitable choice for this application. It ensures that the RL algorithm is evaluated not only for its long-term performance but also for its ability to faithfully replicate the expert’s short-term decision-making, which is often essential for real-world reliability and user trust. \qed 
	\end{example}

	We conclude this section by showing that the CK distance is a \#P-hard function. 
	\begin{theorem} \label{thm:exact-sharp-p} 
		The function $d$ as defined in Definition~\ref{def:ck} is \#P-hard. 
	\end{theorem}

	Theorem~\ref{thm:exact-sharp-p} implies that, unless P = NP, there is no algorithm that computes exactly the CK distance in polynomial time.

	\begin{remark} \label{rem:open-q-computable}
		Although we proved that the CK distance is \#P-hard, the question of whether the problem of computing exactly the CK distance is decidable or not is still open. However, as we will show in Proposition~\ref{sec:approx}, the CK distance can be computed up to an arbitrarily small precision, making it computable in that sense. \qed 
	\end{remark}

	\section{Continuity Properties} \label{sec:continuity}

	In this section, we study \emph{continuity properties} of the CK distance. In particular, we will study \emph{parameter continuity} and a form of ``linear-time continuity'' known as $\Phi$-continuity \cite{Jaeger2014}, defined hereinafter. 

	\begin{definition}[Parameter continuity] \label{def:parameter-cont}
		A distance $d$ between LMCs is said to be \emph{parameter} continuous if, for all $\varepsilon > 0$, there exists $\delta > 0$ such that, for any two LMCs $\Sigma_1$ and $\Sigma_2$ defined on the same set of labels, 
		\[  
			\text{$\Sigma_1$ and $\Sigma_2$ are $\delta$-approximately bisimilar} \implies d(\Sigma_1, \Sigma_2) \leq \varepsilon. 
		\]
	\end{definition}

	\begin{definition}[$\Phi$-continuity]
		Let $\phi \subseteq A^\omega$ be a property. A distance $d$ between LMCs is said to be \emph{$\phi$-continuous} if, for all $\varepsilon > 0$, there exists $\delta > 0$ such that, for any two LMCs $\Sigma_1$ and $\Sigma_2$ defined on the same set of labels, 
		\[
			d(\Sigma_1, \Sigma_2) \leq \delta \implies \sup_{w \in \phi} |p^{\omega}_1(w) - p^{\omega}_2(w)| \leq \varepsilon. 
		\]
		Let $\Phi \subseteq 2^{A^\omega}$ be a class of properties, a distance $d$ is said to be \emph{$\Phi$-continuous} if it is $\phi$-continuous for all $\phi \in \Phi$. 
	\end{definition}

	\begin{remark} \label{rem:parameter-cont}
		Whilst it shares the same spirit, note that the notion of parameter continuity as defined in this paper is less restrictive that the notion of parameter continuity in \cite[Definition~5]{Jaeger2014}. \qed 
	\end{remark}

	Dovetailing on Remark~\ref{rem:jaeger}, Jaeger et al. already showed in \cite{Jaeger2014} that this distance is both parameter and (B-LTL)-continuous\footnote{By abuse of notation, (B-LTL)-continuity refers to $\Phi_{\text{B-LTL}}$-continuity, where $\Phi_{\text{B-LTL}}$ is defined in Section~\ref{sec:language}, as defined above}. In this section, we extend the results in \cite{Jaeger2014} in two separate ways: 
	\begin{enumerate}
		\item We quantify in Proposition~\ref{prop:ck-bound} the value of $\varepsilon$ such that
		\[ 
			\text{$\Sigma_1$ and $\Sigma_2$ are $\delta$-approximately bisimilar} \implies d(\Sigma_1, \Sigma_2) \leq \varepsilon. 
		\]
		\item We quantify in Proposition~\ref{prop:safety} the value of $\varepsilon$ such that
		\[ 
			d(\Sigma_1, \Sigma_2) \leq \delta \implies |p_1^k(w) - p_2^k(w)| \leq \varepsilon, 
		\] 
		for all finite words $w \in A^k$. 
	\end{enumerate} 
	Hence, we recover the less general continuity results shown in \cite{Jaeger2014}, as corollaries of these two propositions (see Corollary~\ref{cor:parameter-cont} and Corollary~\ref{cor:bltl-cont}).

	We start by showing that, if two LMCs are $\delta$-approximately bisimilar, then the CK distance is upper bounded by a function of $\delta$. To prove this result, we rely on a result proven by Bian and Abate in \cite[Theorem~4]{Bian2017}.
	\begin{proposition}[\cite{Bian2017}] \label{prop:tv-bound}
		Given $\delta \in (0, 1)$, let $\Sigma_1$ and $\Sigma_2$ be two $\delta$-approximately bisimilar LMCs defined on the same set of labels. For all $k \geq 0$, it holds that 
		\begin{equation} \label{eq:tv-bound}
			\mathsf{TV}(p_1^k, p_2^k) \leq 1 - (1-\delta)^k, 
		\end{equation}
		where $p_1^k$ and $p_2^k$ are the distributions on $A^k$ respectively generated by $\Sigma_1$ and $\Sigma_2$. 
	\end{proposition}
	For the sake of conciseness, we skip the proof of Proposition~\ref{prop:tv-bound}, as it is similar to that in \cite{Bian2017}. We now provide a bound on the CK distance.
	\begin{proposition} \label{prop:ck-bound}
		Given $\delta \in (0, 1)$, let $\Sigma_1$ and $\Sigma_2$ be two $\delta$-approximately bisimilar LMCs defined on the same set of labels. It holds that 
		\begin{equation} \label{eq:ck-bound}
		\begin{aligned}
			d(\Sigma_1, \Sigma_2) \leq \frac{m\delta}{m - 1 + \delta}. 
		\end{aligned}
		\end{equation}
	\end{proposition}

\begin{figure}
		\centering
		\includegraphics[width = 0.8\textwidth]{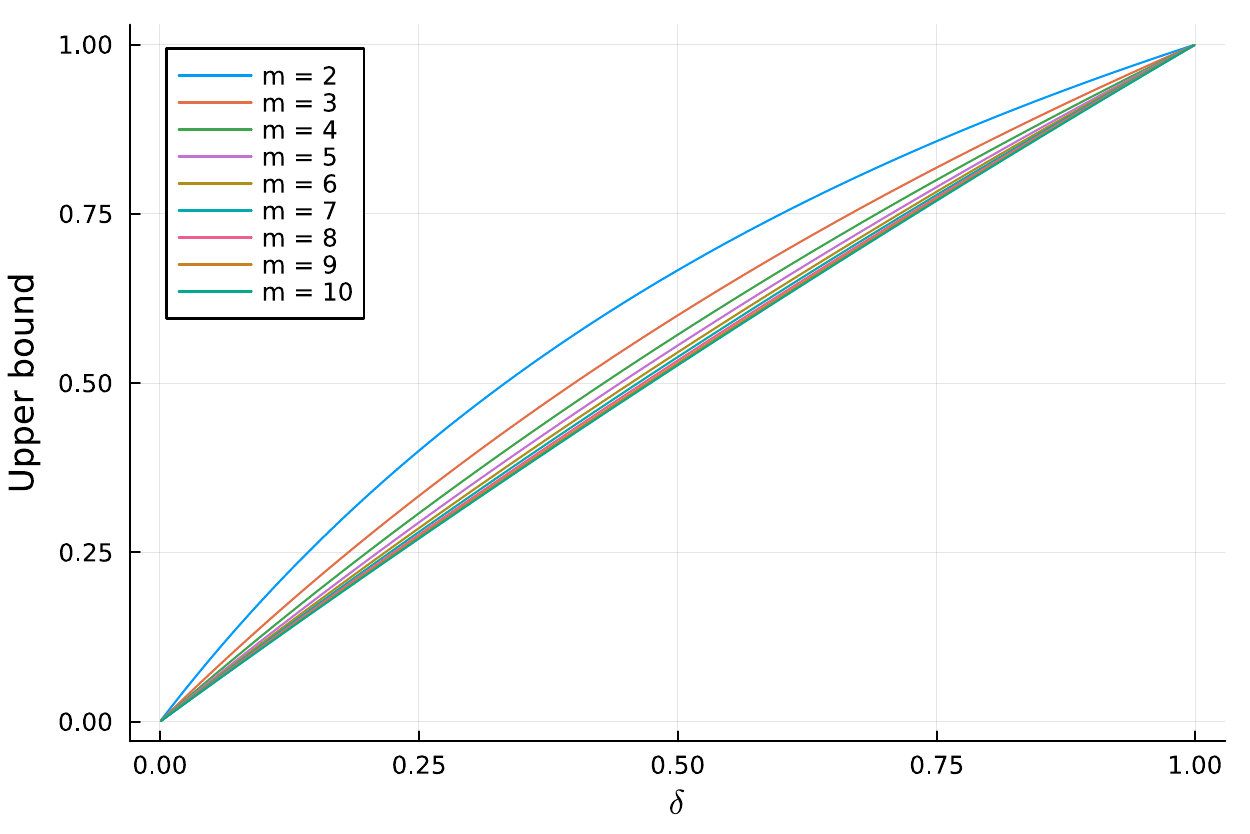}
		\caption{Upper bound \eqref{eq:ck-bound} on the CK distance between two $\delta$-approximately bisimilar LMCs. The upper bound is given as a function $\delta \in (0, 1)$, and for different number of labels.}
		\label{fig:ck-bound}
	\end{figure}

	An illustration of the bound \eqref{eq:ck-bound} can be found in Figure~\ref{fig:ck-bound} for the cases $m = 2, \dots, 10$. A second illustration is also presented in Section~\ref{sec:approx} (see Figure~\ref{fig:Sk}). 

	Proposition~\ref{prop:ck-bound} provides practical insight about the CK distance. Suppose that a lower bound $d(\Sigma_1, \Sigma_2) \geq \underline{d}$ is known. Then, let 
	\[ 
		\overline{\delta} := (m-1)\underline{d} / (m-\underline{d}), 
	\]
	it is impossible that $\Sigma_1$ and $\Sigma_2$ are $\delta$-approximately bisimilar for all $\delta \leq \overline{\delta}$. Proposition~\ref{prop:ck-bound} also provides theoretical insight, as it offers a proof for a less restrictive notion of parameter continuity (see Remark~\ref{rem:parameter-cont}).
	\begin{corollary} \label{cor:parameter-cont}
		The CK distance is parameter continuous in the sense of Definition~\ref{def:parameter-cont}. 
	\end{corollary}

	We now show that, if $d(\Sigma_1, \Sigma_2) \leq \delta$, then the TV distance $\mathsf{TV}(p_1^k, p_2^k)$ is upper bounded by a function of $\delta$ and the horizon $k$. 
	\begin{proposition} \label{prop:safety}
		Let $\Sigma_1$ and $\Sigma_2$ be two LMCs defined on the same set of labels. For all $k \geq 1$, it holds that, if $d(\Sigma_1, \Sigma_2) \leq \delta$, then $\mathsf{TV}(p_1^k, p_2^k) \leq m^{k-1}  \delta$. 
	\end{proposition}

	Again, Proposition~\ref{prop:safety} provides a practical insight about the CK distance. Suppose that an upper bound $d(\Sigma_1, \Sigma_2) \leq \overline{d}$ is known. Given a threshold $\varepsilon \in (0, 1)$, for all 
	\[ 
		k \leq 1 + \left\lfloor \frac{\log(\varepsilon / \overline{d})}{\log(m)} \right\rfloor, 
	\] 
	then $\mathsf{TV}(p_1^k, p_2^k) \leq \varepsilon$. Therefore, given a threshold $\varepsilon$ and an upper bound $d(\Sigma_1, \Sigma_2) \leq \overline{d}$, the result above gives a maximal horizon $k$ such that $|p_1^k(w) - p_2^k(w)| \leq \varepsilon$ for all $k$-long word $w \in A^k$.

	Proposition~\ref{prop:safety} also provides theoretical insights as it provides an alternative proof of (B-LTL)-continuity (see proof in Appendix~\ref{app:proof-bltl}). 
	\begin{corollary} \label{cor:bltl-cont}
		The CK distance is (B-LTL)-continuous. 
	\end{corollary}

	\section{Approximation of CK Distance} \label{sec:approx}

	In this section, we provide a practical way to approximate the CK distance by computing the truncated sum 
	\begin{equation} \label{eq:truncated} 
		S_k(\Sigma_1, \Sigma_2) := \sum_{i = 1}^k \left(\frac{m-1}{m^i}\right) \mathsf{TV}(p_1^i, p_2^i)
	\end{equation}
	up to a certain finite horizon $k > 0$.

	\begin{proposition} \label{prop:approx}
		Let $\Sigma_1$ and $\Sigma_2$ be two LMCs defined on the same set of labels. It holds that 
		\begin{equation} \label{eq:approx}
			0 \leq d(\Sigma_1, \Sigma_2) - S_k(\Sigma_1, \Sigma_2) \leq m^{-k}, 
		\end{equation}
		where $S_k$ is defined in \eqref{eq:truncated}. 
	\end{proposition}

	\begin{example} 
		Let $\Sigma$ be the unbiased Markov chain from Example~\ref{example:biased-mc}, and consider biased Markov chains $\tilde{\Sigma}_{\varepsilon}$ with $\varepsilon \in \{10^{-1}, 10^{-2}, 10^{-3}, 10^{-4}\}$. In Figure~\ref{fig:Sk}, we approximated $d(\Sigma, \tilde{\Sigma}_\varepsilon)$ with $S_k(\Sigma_1, \Sigma_2)$ for $k = 1, \dots, 15$, thereby achieving a precision of $m^{-k} = 2^{-15} \approx 3 \cdot 10^{-5}$. \qed 

		\begin{figure}
			\centering
			\includegraphics[width = 0.8\textwidth]{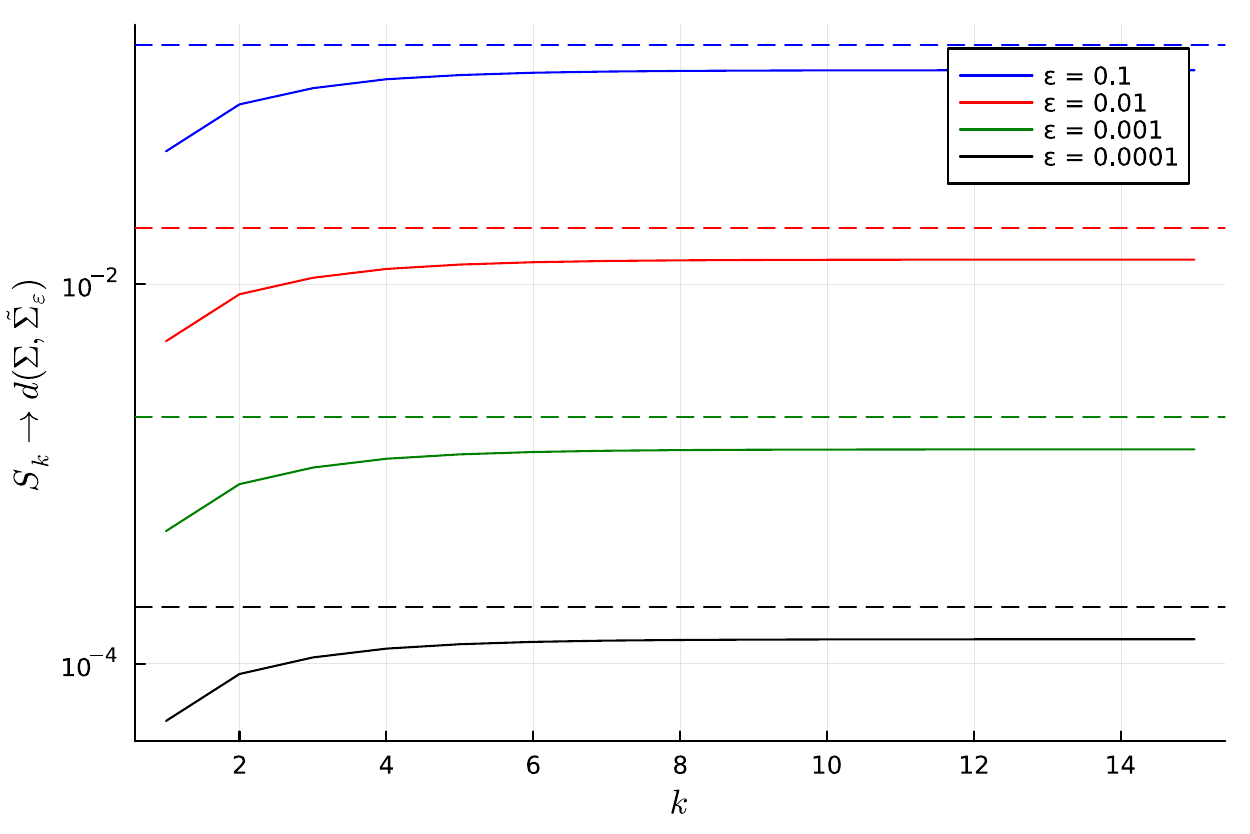}
			\caption{Approximation of the CK distance $d(\Sigma, \tilde{\Sigma}_{\varepsilon})$. The plain line is $S_k(\Sigma_1, \Sigma_2)$, and the dashed line is the upper bound \eqref{eq:ck-bound}. As expected from \eqref{eq:approx}, $S_k(\Sigma_1, \Sigma_2)$ converges.}
			\label{fig:Sk}
		\end{figure}
	\end{example}

	Proposition~\ref{prop:approx} implies that, in order to reach an error of at most $\epsilon \in (0, 1)$, it suffices to choose
	\[
		k \geq \left\lceil \frac{\log(\epsilon^{-1})}{\log(m)} \right\rceil.
	\] 

	Following up on the end of Section~\ref{sec:ck}, we show that the function $S_k$ as described above is \#P-hard, similarly as the exact CK distance. 
	\begin{proposition} \label{prop:approx-is-hard}
		The function $S_k$ as defined in \eqref{eq:truncated} is \#P-hard. 
	\end{proposition}

	Proposition~\ref{prop:approx-is-hard} shows that, unless P = NP, there exists no polynomial algorithm to compute $S_k(\Sigma_1, \Sigma_2)$. In \cite{Banse2025}, an algorithm to compute $S_k(\Sigma_1, \Sigma_2)$ in $k(|S_1|^2 + |S_2|^2) + \mathcal{O}(m^{k+1})$ iterations is presented. 

	\section{Conclusions and Open Questions} \label{sec:conclu}

	In this work, we extended the theoretical foundations of the Cantor-Kantorovich distance. More precisely, building on the central result that the CK distance is a sum of discounted TV distances, we proved that the exact computation of the latter is \#P-hard. We provided an upper bound on the CK distance, and showed that a bounded CK distance $d(\Sigma, \tilde{\Sigma})$ implies a bounded error between $\tilde{\Sigma}$ and $\Sigma$ over finite horizons. Finally, we provided a computable approximation scheme, but showed that the latter is also \#P-hard 

	Some questions remain open for further work. For example, we do not know whether the CK distance is computable. Moreover, even though a specific approximation scheme was presented in this paper, the question of whether there exists a polynomial algorithm for alternatively approximating the CK distance is still open.  

	%
	%
	%
	\bibliographystyle{abbrv}
	\bibliography{ref}

	\appendix

	\section{Proof of Theorem~\ref{thm:ck-is-tv}}

	In order to prove Theorem~\ref{thm:ck-is-tv}, we will need the two following lemmas. 
	\begin{lemma} \label{lemma:thm-in-tac}
		For a fixed $k \in \mathbb{N}$, for all $i \leq k$, let $p_1^i$ and $p_2^i$ be the $i$-long distributions generated by two LMCs defined on the same set of labels. Let 
		\begin{equation} \label{eq:sum_min}
			M_i = \sum_{w \in A^i} \min\{p_1^i(w), p_2^i(w)\}.
		\end{equation}
		It holds that
		$
			\mathsf{K}_\mathsf{C}(p_1^k, p_2^k)
			= 
			1 - M_1 + \sum_{i = 1}^{k-1} m^{-i} (M_i - M_{i+1})
		$.
	\end{lemma}
	\begin{lemma}[\cite{Chen2014}] \label{lemma:kiefer}
		For a fixed $k \in \mathbb{N}$, let $p_1^k$ and $p_2^k$ be the distributions generated by two LMCs defined on the same set of labels. It holds that 
		\[
			\mathsf{TV}(p_1^k, p_2^k) = 1 - M_k, 
		\]
		where $M_k$ is defined in \eqref{eq:sum_min}.
	\end{lemma}
	Lemma~\ref{lemma:thm-in-tac} is the same result as \cite[Theorem~1]{Banse2025} in a slightly different setting (see Remark~\ref{rem:cantor-is-different}). We skip its proof because it is very similar to that in \cite{Banse2025}. Lemma~\ref{lemma:kiefer} was proven by Chen and Kiefer in \cite{Chen2014}.
	\begin{proof}[of Theorem~\ref{thm:ck-is-tv}]
		First, from Lemma~\ref{lemma:thm-in-tac}, it holds that 
		\[
			\mathsf{K}_\mathsf{C}(p^k_1, p^k_2) = 1 + \sum_{i = 1}^{k-1} \left( \frac{1 - m}{m^i} \right) M_i - m^{-(k-1)}M_k. 
		\]
		Now, replacing $M_i = 1 - \mathsf{TV}(p^i_1, p^i_2)$ from Lemma~\ref{lemma:kiefer} yields 
		\begin{equation*}
		\begin{aligned}
			\mathsf{K}_\mathsf{C}(p^k_1, p^k_2) &= 1 + (1 - m) \sum_{k-1} m^{-i} - m^{-(k-1)} 
			\\&\quad\quad + \sum_{i = 1}^{k-1}\left(\frac{m-1}{m^i}\right) \mathsf{TV}(p^i_1, p^i_2) + m^{1-k}\mathsf{TV}(p_1^k, p_2^k). 
		\end{aligned}
		\end{equation*}
		It remains to see that the first three terms of the equation above cancel, which implies 
		\[
			\mathsf{K}_\mathsf{C}(p^k_1, p^k_2) =  \sum_{i = 1}^{k-1}\left(\frac{m-1}{m^i}\right) \mathsf{TV}(p^i_1, p^i_2) + m^{1-k}\mathsf{TV}(p_1^k, p_2^k). 
		\]
		Since $\mathsf{TV}(p_1^k, p_2^k)$ converges to a finite value and that $\lim_{k \to \infty} m^{1-k} = 0$, taking the equation to the limit $k \to \infty$ yields the statement, and the proof is concluded. 
	\end{proof}

	\section{Proof of Corollary~\ref{cor:ck-bounds}}
	\begin{proof}
		For the sake of conciseness, we write $S_k := S_k(\Sigma_1, \Sigma_2)$ all along the proof. By Theorem~\ref{thm:ck-is-tv}, it suffices to show that the sequence of partial sums $(S_k)_{k \geq 0}$ converges, where 
		\[
			S_k = \sum_{i = 1}^k \left( \frac{m-1}{m^i} \right) \mathsf{TV}(p^i, q^i).
		\]
		Let 
		\[
			\overline{S}_k = \sum_{i = 1}^k \left( \frac{m-1}{m^i} \right) = 1 - \frac{1}{m^k}. 
		\]
		Since $0 \leq \mathsf{TV}(p^i, q^i) \leq 1$ for all $i$, it holds that $0 \leq S_k \leq \overline{S}_k$ for all $k$. By the comparaison test, $(S_k)_{k \geq 0}$ converges to a value that is lower bounded by $0$ and upper bounded by $\lim_{k \to \infty} \overline{S}_k = 1$.  
	\end{proof}

	\section{Proof of Theorem~\ref{thm:exact-sharp-p}}

	To prove the latter, we rely on the following result proven by Bhattacharyya et al. in \cite{Bhattacharyya2023}. 
	\begin{definition}[Product distribution] \label{def:product-dis}
		For any $k \in \mathbb{N}_{> 0}$, a \emph{product distribution} $p$ over $\{0, 1\}^k$ is succintly described by $k$ parameters $p_1, \dots, p_k$, where $p_i \in [0, 1]$ is independently
			the probability that the $i$-th coordinate equals 1.    
	\end{definition}
	\begin{proposition}[\cite{Bhattacharyya2023}]
		The TV distance between any two product distributions is \#P-complete. 
	\end{proposition}

	In the following, we will show that, for any two product distributions $p^k, q^k$ over $\{0, 1\}^k$, the computation of $\mathsf{TV}(p^k, q^k)$ can be reduced to the computation of the CK distance in polynomial time. 

	\begin{proof}[of Theorem~\ref{thm:exact-sharp-p}]
		First, given a product distribution $p^k$ described by $k$ parameters $p_1, \dots, p_k$, we introduce a LMC $\Sigma_{p^k}$ that \emph{encodes} $p^k$. Let $\Sigma_{p^k} = (S, \{0, 1\}, \mu, P, L)$ be such that 
		\begin{itemize}
			\item $S$ is a set of $2k$ states $\mathbf{0}_1, \dots, \mathbf{0}_k$ and $\mathbf{1}_1, \dots, \mathbf{1}_k$, 
			\item $\mu$ is such that $\mu(\mathbf{1}_1) = p_1$, $\mu(\mathbf{0}_1) = 1 - p_1$, and $\mu(\mathbf{1}_i) = \mu(\mathbf{0}_i) = 0$ for all $i = 2, \dots, k$, 
			\item $P$ is such that, for all $i = 2, \dots, k$, $P(\mathbf{1}_{i-1}, \mathbf{1}_i) = p_i$, $P(\mathbf{1}_{i-1}, \mathbf{0}_i) = 1 - p_i$, $P(\mathbf{0}_{i-1}, \mathbf{1}_i) = p_i$, $P(\mathbf{0}_{i-1}, \mathbf{0}_i) = 1 - p_i$, and all other entries are null. Also, we require that $P(\mathbf{0}_k, \mathbf{0}_k) = P(\mathbf{1}_k, \mathbf{0}_k) = 1$.
			\item $L$ is such that $L(\mathbf{1}_i) = 1$ and $L(\mathbf{0}_i) = 0$ for all $i = 1, \dots, k$.
		\end{itemize}

		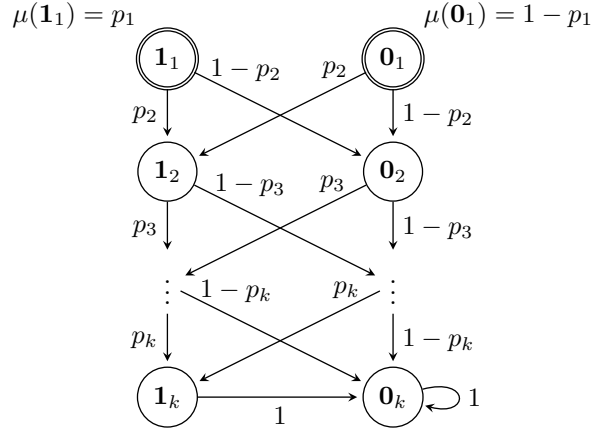
\begin{figure}
			\centering
			\begin{tikzpicture}[->, >=stealth, shorten >= 2pt, line width=0.5pt, node distance=1.5cm]
				\node[circle,draw,double](one1){$\mathbf{1}_1$};
				\node [above left] at (one1.north west) {$\mu(\mathbf{1}_1)= p_1$};
				\node[circle,draw](one2)[below of=one1]{$\mathbf{1}_2$};
				\node[](dots)[below of=one2]{$\vdots$};
				\node[circle,draw](onek)[below of=dots]{$\mathbf{1}_k$};

				\node[circle,draw,double](zero1)[right of=one1, xshift=1.5cm]{$\mathbf{0}_1$};
				\node [above right] at (zero1.north east) {$\mu(\mathbf{0}_1)= 1 - p_1$};
				\node[circle,draw](zero2)[below of=zero1]{$\mathbf{0}_2$};
				\node[](dotsp)[below of=zero2]{$\vdots$};
				\node[circle,draw](zerok)[below of=dotsp]{$\mathbf{0}_k$};

				\path(one1) edge node [left] {$p_2$} (one2); 
				\path(one2) edge node [left] {$p_3$} (dots);
				\path(dots) edge node [left] {$p_k$} (onek);   

				\path(zero1) edge node [right] {$1 - p_2$} (zero2); 
				\path(zero2) edge node [right] {$1 - p_3$} (dotsp);
				\path(dotsp) edge node [right] {$1 - p_k$} (zerok);   

				\path(one1) edge node [above,pos=0.3,yshift=1mm] {$1 - p_2$} (zero2);
				\path(one2) edge node [above,pos=0.3,yshift=1mm] {$1 - p_3$} (dotsp); 
				\path(dots) edge node [above,pos=0.3,yshift=1mm] {$1 - p_k$} (zerok); 

				\path(zero1) edge node [above,pos=0.18] {$p_2$} (one2); 
				\path(zero2) edge node [above,pos=0.18] {$p_3$} (dots); 
				\path(dotsp) edge node [above,pos=0.18] {$p_k$} (onek); 

				\path (zerok) edge [loop right] node [right] {$1$} (zerok);
				\path (onek) edge [below] node [below] {$1$} (zerok);
			\end{tikzpicture}
			\caption{The LMC $\Sigma_p$ that encodes the product distribution $p$ described by parameters $p_1, \dots, p_k$.}
			\label{fig:product_lmc}	
		\end{figure}

		An illustration of $\Sigma_p$ is provided in Figure~\ref{fig:product_lmc}. The LMC $\Sigma_{p^k}$ is as follows: 
		\begin{itemize}
			\item All distributions generated by the LMC $\Sigma_{p^k}$ (in the sense of \eqref{eq:prob-generated}) up to horizon $i \leq k$, noted $p^i$, are the product distributions described by the first $i$ parameters $p_1, \dots, p_i$.
			\item And all distributions generated by the LMC $\Sigma_{p^k}$ up to horizon $i > k$, are such that, for $w = (a_1, \dots, a_k, a_{k+1}, \dots, a_i)$, 
			\[ 
				p^i(w) = \begin{cases}
					p^k(a_1, \dots, a_k) &\text{if } (a_{k+1}, \dots, a_i) = (0, \dots, 0) \\
					0 &\text{otherwise.}
				\end{cases} 
			\]
		\end{itemize}

		Now, given two product distributions $p^k$ and $q^k$ respectively described by $p_1, \dots, p_k$ and $q_1, \dots, q_k$, let $\Sigma_{p^k}$ and $\Sigma_{q^k}$ be the two LMCs that encode $p^k$ and $q^k$, respectively. A first observation is that, for all $i > k$, it holds that
		\begin{equation*}
		\begin{aligned}
			\mathsf{TV}(p^i, q^i) &= \frac{1}{2} \sum_{w \in \{0, 1\}^i} |p^i(w) - q^i(w)| \\
			&= \frac{1}{2} \sum_{w' \in \{0, 1\}^k} | p^i(w'0 \dots 0) - q^i(w'0 \dots 0) | \\
			&= \frac{1}{2} \sum_{w' \in \{0, 1\}^k} | p^k(w') - q^k(w') | \\
			&= \mathsf{TV}(p^k, q^k). 
		\end{aligned} 
		\end{equation*}
		The latter yields
		\begin{equation} \label{eq:linear_combi}
		\begin{aligned} 
			d(\Sigma_{p^k}, \Sigma_{q^k}) &= \sum_{i = 1}^\infty 2^{-i} \mathsf{TV}(p^i, q^i) \\
			&= \sum_{i = 1}^{k-1} 2^{-i} \mathsf{TV}(p^i, q^i) + \mathsf{TV}(p^k, q^k) \left(\sum_{i = k}^\infty 2^{-i}\right) \\
			&= \sum_{i = 1}^{k-1} 2^{-i} \mathsf{TV}(p^i, q^i) + 2^{1-k} \mathsf{TV}(p^k, q^k). 
		\end{aligned} 
		\end{equation}

		In order to lighten the notation, for all $i \leq k$, let $d_i := d(\Sigma_{p^i}, \Sigma_{q^i})$ and $\mathsf{TV}_i = \mathsf{TV}(p^i, q^i)$, it holds from \eqref{eq:linear_combi} that
		\[
			\begin{pmatrix}
				d_1 \\
				d_2 \\
				d_3 \\
				\vdots \\
				d_{k-2} \\
				d_{k-1} \\
				d_k
			\end{pmatrix}
			= 
			\begin{pmatrix}
				1  \\
				2^{-1} & 2^{-1}  \\
				2^{-1} & 2^{-2} & 2^{-2}  \\
				\vdots & \vdots & \vdots & \ddots \\
				2^{-1} & 2^{-3} & 2^{-2} & \dots & 2^{3-k}  \\ 
				2^{-1} & 2^{-3} & 2^{-2} & \dots & 2^{2-k} & 2^{2-k} \\
				2^{-1} & 2^{-3} & 2^{-2} & \dots & 2^{2-k} & 2^{1-k} & 2^{1-k}
			\end{pmatrix}
			\begin{pmatrix}
				\mathsf{TV}_1 \\
				\mathsf{TV}_2 \\
				\mathsf{TV}_3 \\
				\vdots \\
				\mathsf{TV}_{k-2} \\
				\mathsf{TV}_{k-1} \\
				\mathsf{TV}_k \\
			\end{pmatrix}
		\] 
		The matrix in the expression above is invertible since it is a triangular matrix whose diagonal entries are all strictly positive. Therefore, one can compute $\mathsf{TV}_k = \mathsf{TV}(p^k, q^k)$ in polynomial time by computing $k$ times the CK distance, and solving the linear system of equations above. By Proposition~\ref{prop:prove-hardness}, this observation concludes the proof since the computation of $\mathsf{TV}(p^k, q^k)$ was reduced to the computation of the CK distance in polynomial time. 
	\end{proof}

	\section{Proof of Proposition~\ref{prop:ck-bound}}

	\begin{proof} 
		By Theorem~\ref{thm:ck-is-tv} and applying the bound \eqref{eq:tv-bound}, it holds that 
		\begin{equation*}
		\begin{aligned}
			d(\Sigma_1, \Sigma_2) &= \sum_{i = 1}^\infty \left(\frac{m-1}{m^i}\right) \mathsf{TV}(p_1^i, p_2^i) \\
			&\leq \sum_{i = 1}^\infty \left(\frac{m-1}{m^i}\right) (1 - (1-\delta)^i) \\
			&= (m-1) \sum_{i = 1}^\infty  \left(\frac{1}{m}\right)^i - (m-1) \sum_{i = 1}^\infty \left(\frac{1 - \delta}{m}\right)^i  \\
			&= 1 - \frac{(m-1)(1-\delta)}{m-1+\delta} = \frac{m\delta}{m-1+\delta}.
		\end{aligned}
		\end{equation*}
	\end{proof}

	\section{Proof of Corollary~\ref{cor:parameter-cont}} \label{app:proof-pc}
	\begin{proof}
		For all $\varepsilon \in (0, 1)$, let $\delta(\varepsilon) = (m-1)\varepsilon / (m-\varepsilon)$. By Proposition~\ref{prop:ck-bound}, for all $\varepsilon \in (0, 1)$, if $\Sigma_1$ and $\Sigma_2$ are $\delta(\varepsilon)$-approximately bisimilar to each other, then $d(\Sigma_1, \Sigma_2) \leq \varepsilon$, and the proof is completed. 
	\end{proof}

	\section{Proof of Proposition~\ref{prop:safety}}
	\begin{proof}
		Since $\mathsf{TV}(p_1^k, p_2^k)$ is increasing with $k$, by Theorem~\ref{thm:ck-is-tv}, it holds that, for all $k \geq 1$,  
		\begin{equation*}
		\begin{aligned}
			d(\Sigma_1, \Sigma_2) &= \sum_{i = 1}^\infty \left(\frac{m-1}{m^i}\right) \mathsf{TV}(p_1^i, p_2^i) \\
			&\geq \sum_{i = k}^\infty \left(\frac{m-1}{m^i}\right) \mathsf{TV}(p_1^i, p_2^i) \\
			&\geq \mathsf{TV}(p_1^k, p_2^k) \sum_{i = k}^\infty \left(\frac{m-1}{m^i}\right) \\
			&= m^{1-k} \mathsf{TV}(p_1^k, p_2^k). 
		\end{aligned}
		\end{equation*}
		The latter implies that $m^{1-k}\mathsf{TV}(p_1^k, p_2^k) \leq d(\Sigma_1, \Sigma_2) \leq \delta$, which concludes the proof. 
	\end{proof}

	\section{Proof of Corollary~\ref{cor:bltl-cont}} \label{app:proof-bltl}
	\begin{proof}
		Let $\phi \subseteq A^\omega$ be any B-LTL definable property. Following Lemma~\ref{lemma:bltl-finite-union}, there exists a finite number of finite prefixes $W \subseteq A^*$ such that 
		$
			\phi = \bigcup_{w \in W} wA^\omega. 
		$
		This implies that 
		\begin{equation} \label{eq:bound-infinite-finite-words}
			\sup_{w \in \phi} |p_1^\omega(w) - p_2^\omega(w)| \leq \max_{w \in W} |p_1^{k(w)}(w) - p_2^{k(w)}(w)|, 
		\end{equation}
		where $k(w) = |w|$. Let 
		\[
			w_* := \arg\max_{w \in W} |p_1^{k(w)}(w) - p_2^{k(w)}(w)|, 
		\]
		and $k_* := k(w_*)$. Since $W$ is a finite set of words, such maximum exists. Now, for all $\varepsilon \in (0, 1)$, let $\delta(\varepsilon) = (1-m^{-k_*})\varepsilon$. By Proposition~\ref{prop:safety}, if $d(\Sigma_1, \Sigma_2) \leq \delta(\varepsilon)$, then $\mathsf{TV}(p_1^{k_*}, p_2^{k_*}) \leq \varepsilon$. By definition of the TV distance, the latter is an upper bound on \eqref{eq:bound-infinite-finite-words}, and the proof is completed. 
	\end{proof}

	\section{Proof of Proposition~\ref{prop:approx}}
	\begin{proof}
		The fact that the difference is non-negative follows trivially from Theorem~\ref{thm:ck-is-tv} and the fact that $\mathsf{TV}(p, q) \geq 0$ for any two distributions $p, q$. We now prove the upper bound. By Theorem~\ref{thm:ck-is-tv}, and since $\mathsf{TV}(p, q) \leq 1$ for any $p, q$, it holds that
		\begin{equation*}
		\begin{aligned}
			&d(\Sigma_1, \Sigma_2) - \sum_{i = 1}^{k} \left(\frac{m - 1}{m^i}\right) \mathsf{TV}(p^i, q^i) \\
			&\quad \quad= \sum_{i = 1}^{\infty} \left(\frac{m - 1}{m^i}\right) \mathsf{TV}(p^i, q^i) - \sum_{i = 1}^{k} \left(\frac{m - 1}{m^i}\right) \mathsf{TV}(p^i, q^i)\\ 
			&\quad \quad= \sum_{i = k+1}^{\infty} \left(\frac{m - 1}{m^i}\right) \mathsf{TV}(p^i, q^i) \\
			&\quad \quad\leq (m - 1) \sum_{i = k+1}^\infty m^{-i} \\
			&\quad \quad= (m - 1) \frac{m^{-(k+1)}}{1 - m^{-1}} \\
			&\quad \quad= m^{-k}, 
		\end{aligned} 
		\end{equation*}
		which concludes the proof.
	\end{proof}

	\section{Proof of Proposition~\ref{prop:approx-is-hard}}
	\begin{proof}
		We prove this result in the same way as for the exact computation of $d(\Sigma_1, \Sigma_2)$ (see Theorem~\ref{thm:exact-sharp-p}): for any two product distributions $p^k$ and $q^k$ over $\{0, 1\}^k$ (as defined in Definition~\ref{def:product-dis}), we show that the computation of $\mathsf{TV}(p^k, q^k)$ can be reduced to the computation of $S_k(\Sigma_1, \Sigma_2)$ in polynomial time. 

		Given two product distributions $p^k$, $q^k$, let $\Sigma_{p^k}$ and $\Sigma_{q^k}$ be the two LMCs that encode $p^k$ and $q^k$ respectively, in the same fashion as in the proof of Theorem~\ref{thm:exact-sharp-p}. Let $S_{k-1}$ and $S_{k}$ be the truncated sum as defined in \eqref{eq:truncated} with the distributions generated by $\Sigma_{p^k}$ and $\Sigma_{q^k}$, respectively up to horizons $k-1$ and $k$, it holds that 
		\[
			\mathsf{TV}(p^k, q^k) = m^k(S_k(\Sigma_1, \Sigma_2) - S_{k-1}(\Sigma_1, \Sigma_2)). 
		\]
		Therefore the computation of $\mathsf{TV}(p^k, q^k)$ is reduced to the computation of $S_k(\Sigma_1, \Sigma_2)$ in polynomial time, and the proof is completed. 
	\end{proof}
\end{document}